 \documentclass[prl,aps,twocolumn,psfig,preprintnumbers,amsmath,amssymb]{revtex4}
\usepackage{graphicx}\usepackage{epsfig}
\usepackage{amsmath}\usepackage{amssymb}\usepackage{amsfonts}
\usepackage{bm}
\usepackage{exscale}
\usepackage{dcolumn}    
\usepackage{color}
\usepackage{soul}
\usepackage{ulem}
\begin{document}
%
%
\title{On the collectivity of Pygmy Dipole Resonance within schematic TDA and RPA models}
%
%
\author{V. Baran$^{1}$}
\author{D.I. Palade$^{1}$}
\author{M. Colonna$^{2}$}
\author{M. Di Toro$^{3}$}
\author{A. Croitoru$^{1}$}
\author{A.I. Nicolin$^{1,4}$}
\affiliation{$^{1}$Faculty of Physics, University of Bucharest, Atomistilor 405, Magurele, Romania}
\affiliation{$^{2}$INFN-LNS, Laboratori Nazionali del Sud, 95123 Catania, Italy}
\affiliation{$^{3}$Physics and Astronomy Dept., University of Catania, Italy}
\affiliation{$^{4}$Horia Hulubei National Institute for Physics and Nuclear Engineering, Reactorului 30, Magurele, Romania}

%
%
%
\begin{abstract}
Within schematic models based on the Tamm-Dancoff Approximation and the Random-Phase Approximation with separable interactions, we investigate the physical conditions which may determine the emergence of the Pygmy Dipole Resonance in the E1 response of atomic nuclei. We find that if some particle-hole excitations manifest a weaker residual interaction, an additional mode will appear, with an energy centroid closer to the distance between two major shells and therefore well below the Giant Dipole Resonance (GDR). This state, together with the GDR, exhausts all the transition strength in the Tamm-Dancoff Approximation and all the Energy Weighted Sum Rule in the Random-Phase Approximation. 
Thus, within our scheme, this mode, which could be associated with the Pygmy Dipole Resonance, is of collective nature. 
By relating the coupling constants appearing in the separable interaction to the
symmetry energy value at and below saturation density we explore 
the role of density dependence of the symmetry energy on the low energy dipole response. 

\end{abstract}
%
%
%
%
%
\maketitle
%
%
%
%
%

In spite of their apparent simplicity, schematic physics models are always very insightful as they provide in a transparent way the essential physical content which determines a specific feature that is shaping an otherwise complex phenomenon. A quite successful class of such models is that devoted to explain within a quantum many-body treatment the emergence of the collective behavior in various microscopic systems \cite{fanRMP1992}, with special emphasis on atomic nuclei \cite{briNPA1957,bro1964}. To this end, it was pointed out that in the presence of a separable residual particle-hole interaction \cite{broPRL1959,broNPA1961} a coherent superposition of one particle - one hole states is generated, which carries almost all the transition strength and is pushed up or down in energy from the unperturbed value.  

The collectivity of the Giant Dipole Resonance (GDR), one of the most robust modes observed in all nuclei \cite{har2001}, is very well captured in such descriptions \cite{balJETP1962,balNPA1963,greRPP1965}. As a consequence of the repulsive particle-hole residual interaction, the energy peak gets closer to the empirical mass parametrization, $E_{GDR}=80A^{-1/3}$, 
at almost twice the value associated with the distance between two major shells $\hbar \omega_0=41A^{-1/3}$. In recent years experimental investigations \cite{aumPS2013,savPPNP2013} evidenced the presence of a resonance-shaped state \cite{adrPRL2005,wiePRL2009,rosPRL2013} below the GDR response but close to the particle threshold energy, exhausting only few percentages of the dipole Energy Weighted Sum Rule (EWSR). The nature of this state is one of the most important open questions in the field and a subject of intense debate \cite{paaRPP2007,paaJPG2010}, with current interpretations spanning 
from a doorway state \cite{laneAP1971} or single-particle E1 strength that fails to join the GDR \cite{gyaPLB1974,cseNPA1978}, to a collective manifestation of some excess neutrons which oscillate against the more stable core \cite{chamPRC1994}. It is then natural to ask if schematic models as those mentioned above are able to provide additional insight about the physical nature of the low-energy dipole response, the role of the symmetry energy and contribute to the interpretation of the experimentally observed features, such as the energy centroid or the EWSR. 

The purpose of this Letter is to investigate the emergence of new exotic modes in neutron rich nuclei and the role of density dependence of the symmetry energy within such schematic models. We start with a very brief overview of these approaches and then analyze possible extensions which do not spoil their main advantages and allow for more general conditions.

For a system of fermions which interact through an effective two-body potential within a shell model approach in the absence of ground-state correlations, one usually defines the particle-hole vacuum $|0\rangle$ and the particle (hole) energies associated with the single-particle excitations $\epsilon_p$ ($\epsilon_h$).The unperturbed particle-hole excitation energies are obtained as $\epsilon_i=\epsilon_p-\epsilon_h$, where {\it i} labels the specific particle-hole configuration. 
Expressing the interaction
among quasiparticles in terms of the difference between direct and exchange terms, as 
$A_{ij} = \bar{V}_{ph'hp'}=V_{ph',hp'}-V_{ph',p'h}=\langle ph'|\hat{V}|hp'\rangle-\langle ph'|\hat{V}|p'h\rangle$, 
within the linear approximation of the equations-of-motion method \cite{kuo1984}, we get the  Tamm-Dancoff Approximation (TDA) equations:
\begin{equation}
\sum_j (\epsilon_{i}\delta_{ij}+A_{ij})X_j^{(n)}=E_n X_i^{(n)}.
\label{eqTDA}
\end{equation} 
Together with the normalization condition $\sum_j|X_j^{(n)}|^2=1$, Eq.s(1) determine the energy $E_n$ of the state $|n \rangle =\Omega_{TDA}^{+ (n)}|0 \rangle$, as well as the amplitudes which define the excitation operator:
\begin{equation}
\Omega_{TDA}^{+ (n)}=\sum_{p,h}X_{ph}^{(n)}a_p^{+}a_h.
\end{equation} 
As a next step, the exchange term is neglected and a separable particle-hole interaction  
$A_{ij} = \lambda Q_i Q_j^{*}$ 
is introduced for the direct one. One then arrives to the dispersion relation:
\begin{equation}
\sum_i \frac{|Q_i|^2}{E_n-\epsilon_i}= \frac{1}{\lambda},
\label{dispTDA}
\end{equation}
which can be solved for $E_n$. From a simple graphical analysis one notices that for positive (negative) $\lambda$ 
one of the solutions of Eq.(\ref{dispTDA})
is pushed up (down) in energy with respect to the unperturbed energies. 
This state $|n_c\rangle$ has a collective nature, as it can be easily seen  from equation (\ref{dispTDA}) if the degenerate case $\epsilon_i=\epsilon$ is considered. Indeed, for this situation the energy of the collective state is given by 
$E_{n_c}=\epsilon+\lambda \sum_i |Q_i|^2 $, while for all others (non-collective) states one finds $E_n=\epsilon$. Moreover, the transition probability $|\langle n_c |Q|0\rangle|^2 =\sum_i |Q_i|^2$, {\it{i.e.}}, the collective state exhausts all the energy-independent sum rule, while the transition probability to non collective p-h states $|n \rangle$ cancels,  $\langle n |Q|0\rangle = 0$.

Allowing for correlations in the ground state, the TDA treatment is upgraded to the Random Phase Approximation (RPA). The
amplitudes which appear in the excitation operator
\begin{equation}
\Omega_{RPA}^{+ (n)}=\sum_{p,h}X_{ph}^{(n)}a_p^{+}a_h+Y_{ph}^{(n)}a_h^{+}a_p,
\end{equation} 
and which obey the normalization conditions $\sum_j(|X_j^{(n)}|^2-|Y_j^{(n)}|^2)=1$ are obtained from the RPA equations
\begin{eqnarray}
\epsilon_{i}X_i^{(n)}+\sum_j (A_{ij} X_j^{(n)}+B_{ij}Y_j^{(n)})=E_n X_i^{(n)}, \\
\epsilon_{i}Y_i^{(n)}+\sum_j (B_{ij}^{*} X_j^{(n)}+A_{ij}^{*} Y_j^{(n)})=-E_n Y_i^{(n)},
\label{eqRPA}
\end{eqnarray}
with $B_{ij}=\bar{V}_{pp'hh'}$. The amplitudes $Y_j$ are a measure of ground state correlations and by setting all $Y_j=0$ we recover the TDA equations. For separable particle-hole interactions $A_{ij}=\lambda Q_i Q_j^{*}$ and $B_{ij}=\lambda Q_i Q_j$ we get the dispersion relation:
\begin{equation}
\sum_i \frac{2 \epsilon_i |Q_i|^2}{E_n^2-\epsilon_i^2}= \frac{1}{\lambda},
\label{dispRPA}
\end{equation}
which, unlike the TDA treatment, admits a double set of solutions, $\pm E_n$. In the degenerate limit the collective state $|n_c\rangle$ has the energy
\begin{equation}
E_{n,RPA}^2=\epsilon^2+2 \lambda \epsilon \sum_i |Q_i|^2= \epsilon(2E_{n,TDA} - \epsilon)     
\label{en_RPA}
\end{equation}
A very specific feature of the RPA collective state is that it exhausts the whole EWSR gathered in the unperturbed case, 
{\it{i.e.}} $\displaystyle E_{n,RPA}|\langle n_c | Q | \tilde{0} \rangle|^2=\epsilon \sum_i|Q_i|^2$.
Here $| \tilde{0} \rangle $ denotes the correlated ground-state.
Summing up, the 
residual particle-hole interaction builds up a state which is a coherent sum of the $\displaystyle |ph\rangle$ states. 
For a repulsive interaction ($\lambda > 0$),
this is characterized by an energy which is pushed upwards from the unperturbed value and carries all the 
strength.
   
The expression of the coupling constant $\lambda$ can be obtained from considerations based on the self-consistency between the vibrating potential and the induced density variations \cite{boh1998}. In the case of the GDR this quantity is determined by the isovector component of the nuclear interaction, i.e. by
the potential contribution to the symmetry energy at saturation. In the expression of the energy per nucleon the symmetry
energy $\displaystyle \frac{E_{sym}}{A}$ 
is the quantity connected to the isospin $\displaystyle I=\frac{N-Z}{A}$ degree of freedom, {\it{i.e.}}
$\displaystyle \frac{E}{A}(\rho,I)=\frac{E}{A}(\rho,I=0)+ \frac{E_{sym}}{A}(\rho) I^2$ and contains both a kinetic contribution
associated with Pauli correlations, as well as a potential contribution determined by the nuclear interaction: 
$\displaystyle \frac{E_{sym}}{A}=b_{sym}^{(kin)}+b_{sym}^{(pot)}$ \cite{barPR2005}. Then 
$\displaystyle \lambda=\frac{6 b_{sym}^{(pot)}(\rho_0)}{A \langle r^2 \rangle}$, where $\langle r^2 \rangle$ is the mean square radius of the nucleus considered
and $\displaystyle \rho_0$ is the saturation density. Considering this value for $\lambda$ and accounting for the sum-rules satisfied by the matrix elements $\displaystyle |Q_i|^2$ \cite{rin1980}, the energy centroid and the EWSR exhausted by the GDR were successfully reproduced by the RPA treatment.

{\it TDA treatment for low-lying modes}.
Finite nuclei, however, exhibit a density profile.
Since the symmetry energy decreases with density, one expects a smaller value of the coupling constant for the nucleons located at the surface. 
This is particularly true for neutron-rich nuclei, where several neutrons
are located in a region at quite low density, the neutron skin.  
Analogous arguments were promoted in phenomenological models \cite{mohPRC1971} when three coupled fluids ({\it{i.e.}}, protons, blocked neutrons and excess neutrons) were considered to describe various normal modes in a hydrodynamical 
picture. We shall implement this idea in a schematic approach by relaxing the condition of a unique coupling constant for all particle-hole pairs. Similar generalizations of the separable interaction were proposed also in microscopic approaches in order to include the coupling between normal and threshold states \cite{barzNPA1980} or to study the GDR in fissioning nuclei \cite{reiNPA1978}. 
To this end, we assume that for a subsystem of particle-hole pairs, namely $i,j \le i_c$, the interaction is 
$\displaystyle   A_{ij}=\lambda_1 Q_i Q_j^{*}$, with $\displaystyle \lambda_1=\lambda(\rho_0)$ corresponding to the potential symmetry energy at saturation density, while for the other subsystem, namely $i,j>i_c$, the interaction is characterized by a weaker strength $\displaystyle A_{ij}=\lambda_3 Q_i Q_j^{*}$, with $\lambda_3=\lambda(\rho_e)$ associated with the symmetry energy value 
at a much lower density $\rho_e<<\rho_0$. If $\displaystyle i \le i_c , j > i_c$ or $\displaystyle i > i_c , j \le i_c$, {\it i.e.}, for the coupling between the two subsystems, we consider $\displaystyle A_{ij}=\lambda_2 Q_i Q_j^{*}$ with $\lambda_2=\lambda(\rho_i)$ corresponding to a potential symmetry energy at an intermediate density  $\rho_0>\rho_i>\rho_e$ and consequently $\lambda_1>\lambda_2>\lambda_3>0$. The TDA equations for the corresponding amplitudes $\displaystyle X_i^{(n)}$ can be generalized straightforwardly as
\begin{eqnarray}
\epsilon_i X_i^{(n)}+\lambda_1 Q_i \sum_{j \le i_c}Q_j^{*}X_j^{(n)}+\lambda_2 Q_i \sum_{j >i_c}Q_j^{*}X_j^{(n)} =E_n X_i^{(n)} \nonumber\\
~~~{\mbox{if}}~~~i \le i_c,~~~~~~~  \\
\epsilon_i X_i^{(n)}+\lambda_2 Q_i \sum_{j \le i_c}Q_j^{*}X_j^{(n)}+\lambda_3 Q_i \sum_{j > i_c}Q_j^{*}X_j^{(n)} =E_n X_i^{(n)} \nonumber\\
~~~{\mbox{if}}~~~i > i_c,~~~~~~~  
\end{eqnarray}
with the solutions
\begin{eqnarray}
X_i^{(n)}=\frac{N^{c}}{E_n-\epsilon_i}Q_i~~ if~~~i \le i_c, 
\label{ampTDA1} \\
X_i^{(n)}=\frac{N^{e}}{E_n-\epsilon_i}Q_i~~ if~~~i > i_c.
\label{ampTDA2}
\end{eqnarray}
Here the normalization factors are given by 
\begin{eqnarray}
N^{c}=\lambda_1 \sum_{j \le i_c}Q_j^{*}X_j^{(n)}+\lambda_2 \sum_{j > i_c}Q_j^{*}X_j^{(n)}, 
\label{normTDA1} \\
N^{e}=\lambda_2 \sum_{j \le i_c}Q_j^{*}X_j^{(n)}+\lambda_3 \sum_{j > i_c}Q_j^{*}X_j^{(n)}.
\label{normTDA2}
\end{eqnarray}
Using  equations (\ref{ampTDA1})-(\ref{normTDA2}) we observe that  $\displaystyle N^{c}$ and $\displaystyle N^{e}$ satisfy the homogeneous system
of equations:
\begin{eqnarray}
\Big( \lambda_1 \sum_{i \le i_c} \frac{|Q_i|^2}{E_n-\epsilon_i}-1 \Big) N^{c} + \lambda_2 \sum_{i > i_c} \frac{|Q_i|^2}{E_n-\epsilon_i} N^{e}
=0,\\
\lambda_2 \sum_{i \le i_c} \frac{|Q_i|^2}{E_n-\epsilon_i} N^{c}+\Big( \lambda_3 \sum_{i > i_c} \frac{|Q_i|^2}{E_n-\epsilon_i}-1 \Big) N^{e}
=0.
\end{eqnarray}
If we resume to the degenerate case $\epsilon_i=\epsilon$, with $\alpha=\displaystyle \sum_{i \le i_c} |Q_i|^2$, $\beta=\displaystyle \sum_{i > i_c} |Q_i|^2$, by imposing to have nontrivial solutions, we get:
\begin{equation}
(E_n-\epsilon)^2-(\lambda_1 \alpha + \lambda_3 \beta) (E_n-\epsilon) +(\lambda_1\lambda_3-\lambda_2^2)\alpha\beta=0.
\label{detTDA}
\end{equation}
Then the TDA collective energies are:
\begin{eqnarray}
E_n^{(1)}=\epsilon+\frac{(\lambda_1 \alpha + \lambda_3 \beta)}{2}\Bigg(1+\sqrt{1-\frac{4(\lambda_1\lambda_3-\lambda_2^2)\alpha\beta}
{(\lambda_1 \alpha + \lambda_3 \beta)^2}}\Bigg)~~
\label{enTDA1} \\
E_n^{(2)}=\epsilon+\frac{(\lambda_1 \alpha + \lambda_3 \beta)}{2}\Bigg(1-\sqrt{1-\frac{4(\lambda_1\lambda_3-\lambda_2^2)\alpha\beta}
{(\lambda_1 \alpha + \lambda_3 \beta)^2}}\Bigg).~~
\label{enTDA2}
\end{eqnarray}
It is obvious from the equation (\ref{detTDA}) that by setting $\lambda_1=\lambda_2=\lambda_3=\lambda$ we return to the standard situation with only one collective energy. Simple expressions for $E_n^{(1)}$ and $E_n^{(2)}$ are obtained if we assume that $\lambda_1 \alpha>>\lambda_3 \beta$:  
\begin{eqnarray}
E_n^{(1)}\approx \epsilon+(\lambda_1 \alpha + \lambda_3 \beta), \\
E_n^{(2)}\approx \epsilon+\frac{(\lambda_1\lambda_3-\lambda_2^2)\alpha\beta}
{(\lambda_1 \alpha + \lambda_3 \beta)}. 
\end{eqnarray}
One of the solutions, $E_n^{(1)}$, is nearest to the value associated with the collective mode obtained in the usual TDA approach while  the other one, $E_n^{(2)}$, is much closer to the unperturbed value $\epsilon$. 
The amplitudes $X_i^{(n_1)}$ and $X_i^{(n_2)}$ will define the two operators whose action on the ground state generates the two
collective states $|n_{c,1} \rangle$ and $|n_{c,2} \rangle$. It is interesting to observe that now 
energy independent sum rule is distributed only between these two states, {\it i.e.},
\begin{equation}
|\langle n_{c,1}|Q|0\rangle|^2+|\langle n_{c,2}|Q|0\rangle|^2=\alpha+\beta=\sum_i |Q_i|^2.
\label{transTDA2}
\end{equation}
We therefore conclude {\it that both states manifest the feature expected for a collective behavior}. 
Equation (\ref{transTDA2}) can be easily derived observing that 
\begin{equation}
\langle n_{c,k}|Q|0 \rangle=\sum_i Q_i X_i^{(n_k) *}=\frac{\alpha+x_k\beta}{\sqrt{\alpha+x_k^2\beta}},
\end{equation}
where $k={1,2}$ and $x_k=((E_n^{(k)}-\epsilon)-\lambda_1\alpha)/\lambda_2\beta$. When all coupling constants become equal the transition amplitude of the state with higher energy goes to $(\alpha+\beta)$, as expected, exhausting all the sum rule. 

{\it RPA treatment for low-lying modes.} 
Including the ground state correlations does not change the main conclusions obtained within the TDA treatment. Also in this case we shall find the appearance of a second collective state if the unique coupling constant condition is relaxed. The equations for forward and backward amplitudes become
\begin{eqnarray}
\epsilon_i X_i^{(n)}+\lambda_1 Q_i(\sum_{j \le i_c}Q_j^{*}X_j^{(n)}+\sum_{j \le i_c}Q_j Y_j^{(n)})+~~~~~~~~~~ \nonumber\\
+\lambda_2 Q_i (\sum_{j > i_c}Q_j^{*}X_j^{(n)}+\sum_{j > i_c}Q_j Y_j^{(n)}) =E_n X_i^{(n)}~~~\nonumber \\
\epsilon_i Y_i^{(n)}+\lambda_1 Q_i^{*}(\sum_{j \le i_c}Q_j^{*}X_j^{(n)}+\sum_{j \le i_c}Q_j Y_j^{(n)})+~~~~~~~~~ \nonumber\\
+\lambda_2 Q_i^{*}(\sum_{j>i_c}Q_j^{*}X_j^{(n)}+\sum_{j>i_c}Q_j Y_j^{(n)}) =-E_n Y_i^{(n)}~~~ \nonumber\\
~~~ {\mbox{if}}~~~i \le i_c,~~~~~  \\
\epsilon_i X_i^{(n)}+\lambda_2 Q_i (\sum_{j \le i_c}Q_j^{*}X_j^{(n)}+\sum_{j \le i_c}Q_j Y_j^{(n)})+~~~~~~~~~~ \nonumber\\
\lambda_3 Q_i (\sum_{j>i_c}Q_j^{*}X_j^{(n)}+\sum_{j>i_c}Q_j Y_j^{(n)}) =E_n X_i^{(n)}~~~ \nonumber\\
\epsilon_i Y_i^{(n)}+\lambda_2 Q_i^{*} (\sum_{j \le i_c}Q_j^{*}X_j^{(n)}+\sum_{j \le i_c}Q_j Y_j^{(n)})+~~~~~~~~~~ \nonumber\\
\lambda_3 Q_i^{*} (\sum_{j>i_c}Q_j^{*}X_j^{(n)}+\sum_{j>i_c}Q_j Y_j^{(n)}) =-E_n Y_i^{(n)}~~~ \nonumber\\
~~~{\mbox{if}}~~~i > i_c,~~~~~  
\label{eqRPA}
\end{eqnarray}
with the solutions
\begin{eqnarray}
X_i^{(n)}=\frac{M^{c}}{E_n-\epsilon_i}Q_i~;~Y_i^{(n)}=-\frac{M^{c}}{E_n+\epsilon_i}Q_i^{*}~ {\mbox{if}}~i \le i_c,~  \\
X_i^{(n)}=\frac{M^{e}}{E_n-\epsilon_i}Q_i~;~Y_i^{(n)}=-\frac{M^{e}}{E_n+\epsilon_i}Q_i^{*}~ {\mbox{if}}~i > i_c.~
\label{ampTDA}
\end{eqnarray}
The normalization factors
\begin{eqnarray}
M^{c}=\lambda_1 \sum_{j \le i_c}(Q_j^{*}X_j^{(n)}+Q_j Y_j^{(n)})+~~~~~~~~~~~~~~~~~~ \nonumber\\
~~~~~~~~~~~~~~~~~~~~\lambda_2 \sum_{j > i_c}(Q_j^{*}X_j^{(n)}+Q_j Y_j^{(n)}),~~~ \\
M^{e}=\lambda_2 \sum_{j \le i_c}(Q_j^{*}X_j^{(n)}+Q_j Y_j^{(n)})+~~~~~~~~~~~~~~~~~~  \nonumber\\
~~~~~~~~~~~~~~~~~~~~\lambda_3 \sum_{j > i_c}(Q_j^{*}X_j^{(n)}+Q_j Y_j^{(n)}).~~~
\label{normRPA}
\end{eqnarray}
satisfy the homogeneous system of equations:
\begin{eqnarray}
\Bigg(\lambda_1 \sum_{i \le i_c} \frac{2\epsilon_i |Q_i|^2}{E_n^2-\epsilon_i^2}-1\Bigg)M^{c} + 
\lambda_2 \sum_{i > i_c} \frac{2\epsilon_i|Q_i|^2}{E_n^2-\epsilon_i^2} M^{e}=0,~~\\
\lambda_2 \sum_{i \le i_c} \frac{2\epsilon_i|Q_i|^2}{E_n^2-\epsilon_i^2} M^{c}+
\Bigg(\lambda_3 \sum_{i > i_c} \frac{2\epsilon_i|Q_i|^2}{E_n^2-\epsilon_i^2}-1\Bigg)M^{e}=0.~~
\end{eqnarray}
In the degenerate case, $\epsilon_i=\epsilon$, nontrivial solutions are obtained if
\begin{equation}
(E_n^2-\epsilon^2)^2-2\epsilon(\lambda_1 \alpha + \lambda_3 \beta) (E_n^2-\epsilon^2)+4\epsilon^2(\lambda_1\lambda_3-\lambda_2^2)\alpha\beta=0.
\label{detRPA}
\end{equation}
Then the collective RPA energies are:
\begin{eqnarray}
E_{n,RPA}^{(1) 2}=\epsilon^2+2\epsilon (E_{n,TDA}^{(1)}-\epsilon)=\epsilon(2E_{n,TDA}^{(1)}-\epsilon), \\
E_{n,RPA}^{(2) 2}=\epsilon^2+2\epsilon (E_{n,TDA}^{(2)}-\epsilon)=\epsilon(2E_{n,TDA}^{(2)}-\epsilon), 
\end{eqnarray}
where $\displaystyle E_{n,TDA}^{(1)}$ and $\displaystyle E_{n,TDA}^{(2)}$ are the corresponding energies in the TDA approximation given by
(\ref{enTDA1},\ref{enTDA2}). It is interesting to notice that within the RPA treatment the total EWSR is shared only by 
these two states, i.e.
\begin{equation}
E_{n,RPA}^{(1)}|\langle n_{c,1}|Q|\tilde{0}\rangle|^2+E_{n,RPA}^{(2)}|\langle n_{c,2}|Q|\tilde{0}\rangle|^2
=\sum_i \epsilon |Q_i|^2,
\label{ewsrRPA2}
\end{equation}
therefore both of them manifest a collective nature. The last relation can be easily deduced observing that ($k=1,2$)
\begin{eqnarray}
\langle n_{c,k}|Q|\tilde{0} \rangle=\sum_i (Q_i X_i^{(n_k) *}+Q_i^{*} Y_i^{(n_k) *})= ~~~~~~~~~~~~~\nonumber \\
={\sqrt \frac{\epsilon}{E_{n,RPA}^{(k)}}} \frac{\alpha+z_k\beta}{\sqrt{\alpha+z_k^2\beta}};~z_k=\frac{E_{n,RPA}^{(k) 2}-\epsilon^2}{2\epsilon \lambda_2\beta}-\frac{ \lambda_1\alpha}{ \lambda_2\beta}.~~
\end{eqnarray} 

In the following we apply the predictions of the schematic TDA and RPA models 
to specific nuclear systems, where the appearence of a low-lying strength has been observed in the isovector
dipole response. Thus we associate the low energy state discussed above with the Pygmy Dipole Resonance (PDR).     
We employ the EWSR associated with the isovector dipolar field corresponding to the unperturbed case: 
$\displaystyle m_1=\hbar \omega_0 (\alpha+\beta)=\frac{\hbar^2}{2 m} \frac{NZ}{A}$. The values for 
$\displaystyle \alpha$ and $\displaystyle \beta$ are related to the number of protons ($\displaystyle Z_c$) and neutrons
($\displaystyle N_c$) which belong to core, ($\displaystyle A_c=N_c+Z_c$) and the number of neutrons considered in excess, {\it i.e.} nucleons at much lower density ($\displaystyle N_e$), respectively. We first consider  $\displaystyle N_e$ as a parameter  ($N_e+N_c=N$) but a more precise value can be estimated from arguments based on density distributions of protons and neutrons, as we discuss later. We then obtain 
$\displaystyle \hbar \omega_0 \alpha=\frac{\hbar^2}{2 m} \frac{N_cZ}{A_c}$ and 
$\displaystyle \hbar \omega_0 \beta=\frac{\hbar^2}{2 m} \frac{N_eZ^2}{AA_c}$ \cite{barRJP2012,barPRC2012}. 
Concerning the coupling constants, we observe that in the presence of the dipolar field the charges of protons and neutrons are considered to be $\displaystyle N/A$ and $\displaystyle -Z/A$, respectively. 
Then $\displaystyle \lambda_1=\frac{A^2}{NZ} \frac{10 b_{sym}^{(pot)}(\rho_0)}{A R^2}$, where the nuclear radius is $R=1.2 A^{1/3}$.  Let us first adopt for   $\displaystyle \lambda_3$ a constant value $\displaystyle \lambda_3=0.2 \lambda_1$ which corresponds to the lower density associated with the neutron skin region and investigate the influence of $\lambda_2$  when varied from 
$\displaystyle \lambda_3$ (a weak coupling between the two subsystems)  to $\displaystyle \lambda_1$ (a strong coupling between the two subsystems). 
\begin{figure}
\begin{center}
\includegraphics*[scale=0.33]{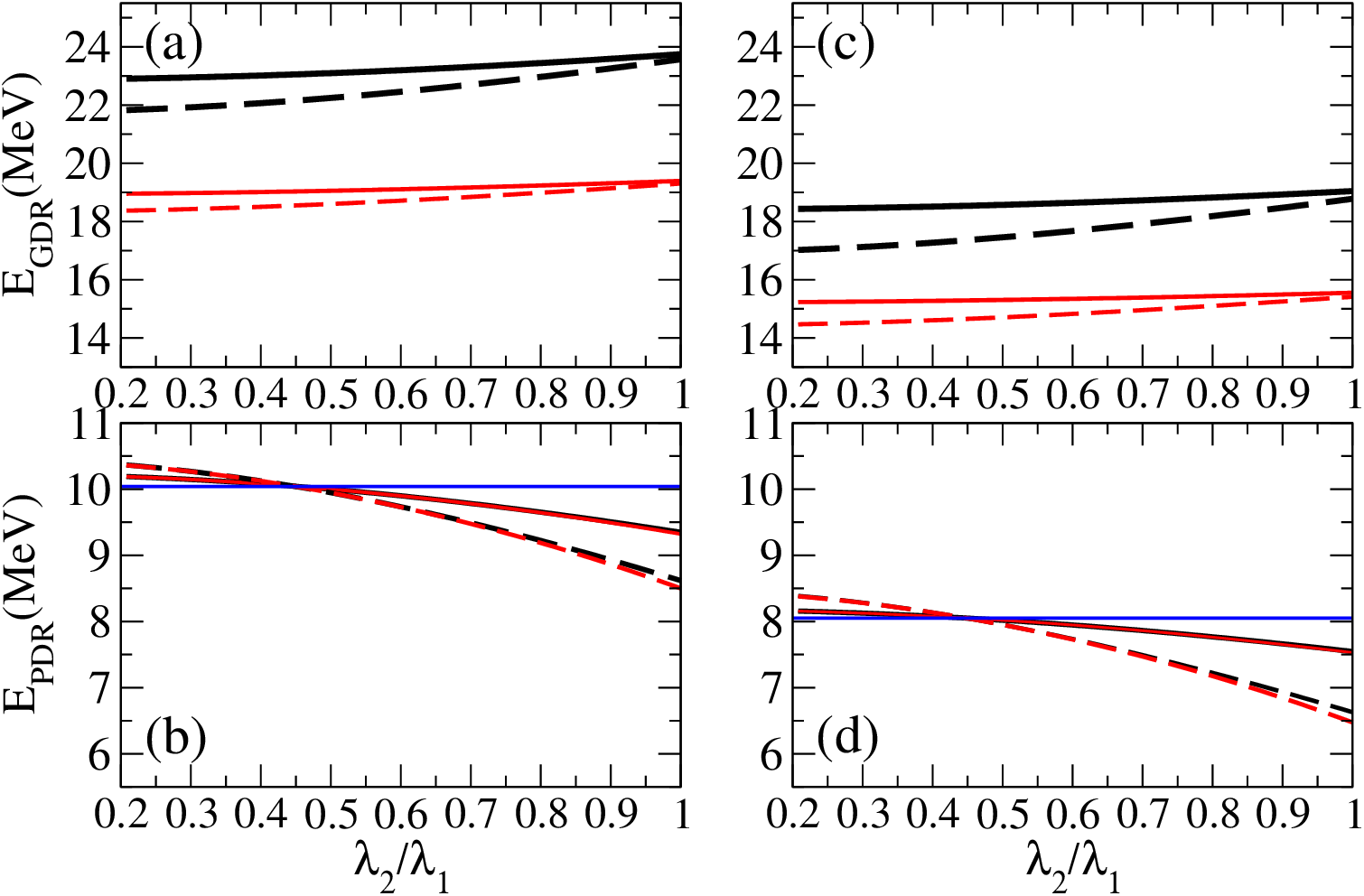}
\end{center}
\caption{(Color online) The GDR and PDR energy centroids as a function of the ratio $\displaystyle \lambda_2/\lambda_1$. 
The black thick lines refer to the TDA while the red lines to RPA calculations. 
For $^{68}$Ni ((a) and (b)) the solid lines correspond to  $\displaystyle N_e=6$; the dashed lines correspond
to  $N_e=12$.
(b) For $^{132}$Sn ((c) and (d)) the solid lines correspond to  $\displaystyle N_e=12$; the dashed lines correspond
to  $\displaystyle N_e=32$. In (b) and (d) the horizontal blue line indicates the unperturbed energy value.}
\label{ener_ni_sn}
\end{figure}
We consider first the nucleus $^{68}$Ni and determine the position of the energy centroids corresponding
to the two collective states both in TDA (black thick lines) and RPA (red lines) calculations, see Figure \ref{ener_ni_sn} (a),(b).
Two values were chosen for the number of excess neutrons, namely $N_e=12$ which corresponds to the extreme case $N_e=N-Z$ (dashed lines) and $N_e=6$ (solid lines). We observe that the ground state correlations are influencing strongly the GDR peak and that the RPA predictions are closer to the experimental values (around 17.8 MeV). The PDR energy centroid does not change much neither when we modify the value of $N_e$, nor when we include the ground state correlations. The experimental value recently reported in \cite{rosPRL2013} is $E_{PDR}^{exp}=9.55$ MeV, while in our study, for $N_e=6$, it changes from  $\displaystyle E_{PDR}=10.2$ MeV to $9.3$ MeV, when $\displaystyle \lambda_2$ increases from
$\displaystyle \lambda_3$ to $\displaystyle \lambda_1$.

 We report the same type of calculations for the $\displaystyle ^{132}$Sn in Figure \ref{ener_ni_sn} (c),(d) considering the cases $\displaystyle N_e=32$ (dashed lines) and $\displaystyle N_e=12$ (solid lines). For this system, when $\displaystyle N_e=12$, the position of the PDR energy centroid changes from $\displaystyle E_{PDR}=8.5$ MeV  to $\displaystyle 7.5$ MeV as
$\displaystyle \lambda_2$ is varied as before. A steeper decrease is observed for a greater value of $\displaystyle N_e$.

\begin{figure}
\begin{center}
\includegraphics*[scale=0.33]{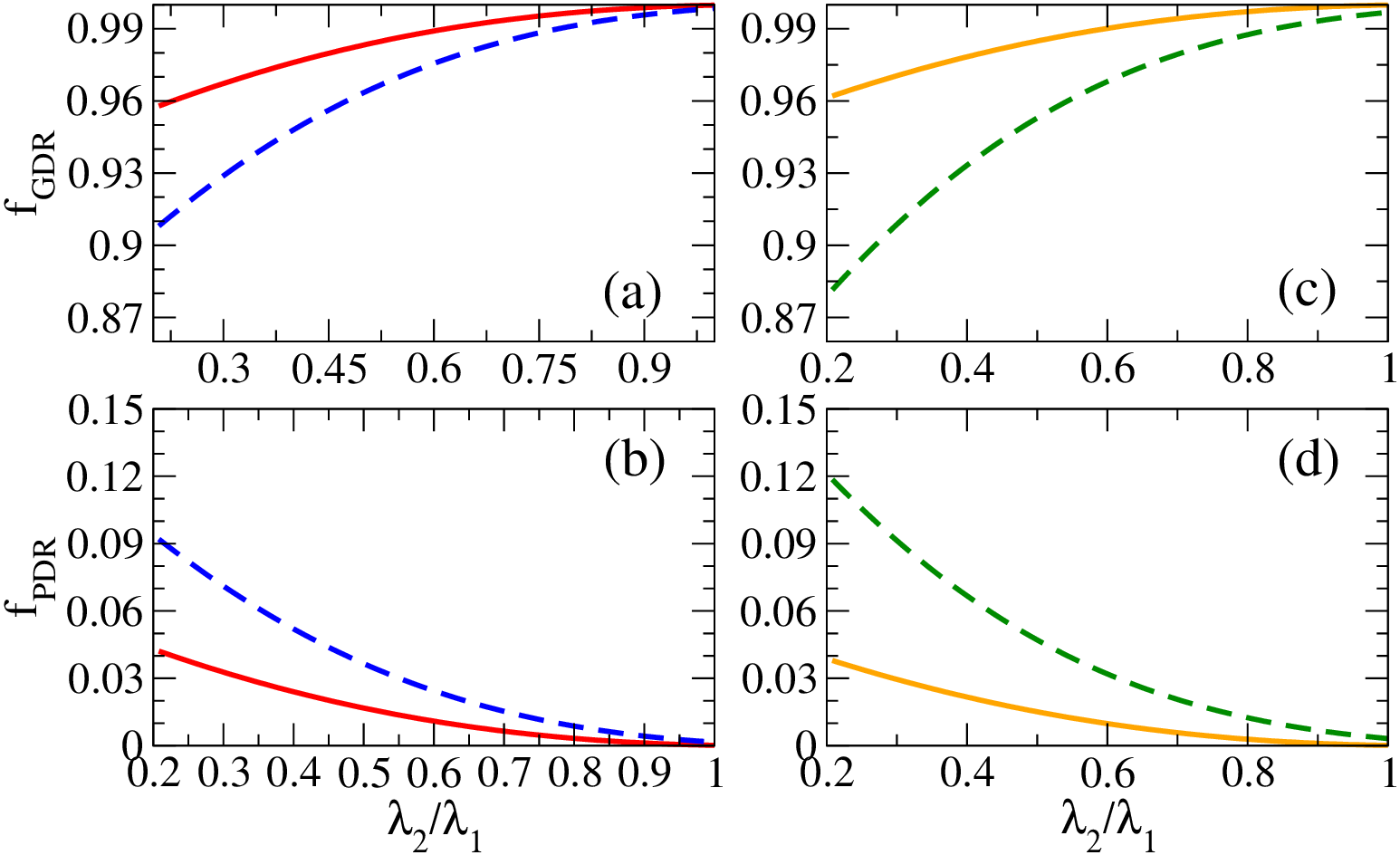}
\end{center}
\caption{(Color online) (a) The EWSR fraction exhausted by GDR in RPA calculations for $^{68}Ni$.
$N_e=6$ (red solid lines) and $\displaystyle N_e=12$ (blue dashed lines).
(b) The EWSR fraction exhausted by PDR in RPA calculations for $^{68}Ni$.
$N_e=6$ (red solid lines) and $\displaystyle N_e=12$ (blue dashed lines).
(c) The EWSR fraction exhausted by GDR in RPA calculations for $^{132}Sn$.
$N_e=6$ (orange solid lines) and $\displaystyle N_e=12$ (blue dashed lines).
(d) The EWSR fraction exhausted by PDR in RPA calculations for $^{132}Sn$.
$N_e=6$ (orange solid lines) and $\displaystyle N_e=12$ (green dashed lines).}
\label{ewsr_ni_sn}
\end{figure}

In Figure \ref{ewsr_ni_sn} we plot the fraction of EWSR exhausted by the GDR ($\displaystyle f_{GDR}$ ) and the PDR ($\displaystyle f_{PDR}$) as predicted by the RPA calculations for the same systems: $\displaystyle ^{68}$Ni, Fig. \ref{ewsr_ni_sn} (a) and (b)  and $^{132}$Sn, Fig. \ref{ewsr_ni_sn} (c) and (d). A greater value of $\displaystyle N_e$ determines a larger value of the EWSR fraction exhausted by the PDR. Moreover, $\displaystyle f_{PDR}$ is strongly influenced by the value of the coupling constant $\displaystyle \lambda_2$ at variance with the $\displaystyle E_{PDR}$ position. 
 In the case of $\displaystyle ^{68}$Ni, for $\displaystyle \lambda_2 / \lambda_1=0.4$, $\displaystyle f_{PDR}$ varies
from $\displaystyle 2.4\%$ to $\displaystyle 5.2\%$ when $\displaystyle N_e$ changes from $6$ to $12$. The experimental values are spanning a domain between $\displaystyle 2.8\%$ and $\displaystyle 5\%$ \cite{wiePRL2009,rosPRL2013}. 

\begin{figure}
\begin{center}
\includegraphics*[scale=0.32]{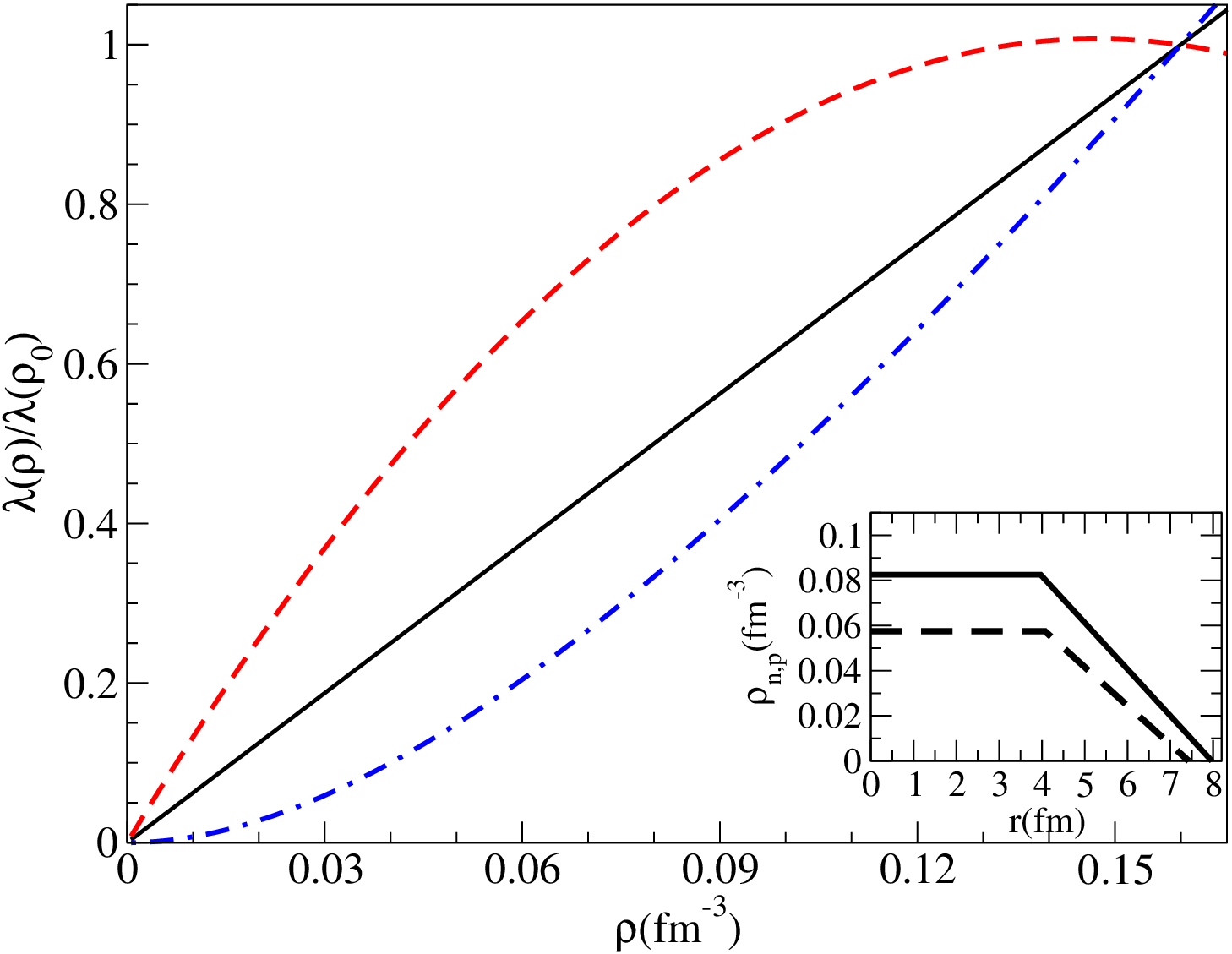}
\end{center}
\caption{(Color online) The ratio $\displaystyle \lambda(\rho)/\lambda(\rho_0)$ as a function of density for 
asystiff EOS (black solid lines), asysuperstiff EOS (blue dot-dashed lines) and
asysoft EOS (red dashed lines). The inset: the trapezoidal distribution of neutron (black solid line) and proton 
(black dashed line) densities for $^{132}Sn$ considered in the calculations.}
\label{lambdarho}
\end{figure}

Our approach also allows an analysis of the role of the symmetry energy when some additional assumptions concerning the connection between the values
of $\displaystyle \lambda_i$ and 
the density behavior of the symmetry energy are established.  Here we employ three
different parameterizations of the potential symmetry energy denoted as asysoft, asystiff and asysuperstiff, respectively \cite{barPR2005}.
The ratio of the coupling constant at a given density $\displaystyle \rho$ to the coupling constant at the saturation density, 
$\displaystyle \lambda(\rho)/\lambda(\rho_0)$ 
is shown in Figure \ref{lambdarho} for the three asy-EOS.  We focus our discussion on $\displaystyle ^{132}Sn$ and approximate the radial proton and
neutron density  distributions by trapezoidal shapes \cite{ikePTP1984}. We reproduce the proton mean-square radius and obtain a neutron skin thickness $\Delta R_{np}=0.3fm$ when we adopt for the central densities the values provided by the Vlasov calculations \cite{barPRC2012}, $\displaystyle \rho_{n}=0.0825 fm^{-3}, \rho_{p}=0.0575 fm^{-3}$, see the inset in Fig. \ref{lambdarho}. We consider the number of neutrons in excess as being determined by the neutron density distribution beyond $r=6.5$ fm, where the tail of the protons distribution is approaching the end part. In this way we obtain a value of $\displaystyle N_e$ around $13.5$ neutrons. We also assume that the average density of these particles will define $\displaystyle \rho_e$, 
obtaining $\displaystyle \rho_e=0.0186fm^{-3}$.
For the three asy-EOS we calculate the corresponding 
$\lambda_3/\lambda_1$ ratio, indicated in Table \ref{taby}. 
\begin{table}
\begin{center}
\begin{tabular}{|l|r|r|r|r|r|r|} \hline
asy-EoS       & $\displaystyle \lambda_2/\lambda_1$ & $\displaystyle \lambda_3/\lambda_1$  & $E_{PDR}$ & $E_{GDR}$ & $\displaystyle f_{PDR}(\%)$ & $\displaystyle f_{PDR}^{V}(\%)$ \\ \hline
asysoft       &     0.57              & 0.23     & 7.98  & 15.30   & 1.3  & 2.4   \\ \hline
asystiff      &     0.31              & 0.11     & 8.05  & 15.20   & 3.3  & 4.2  \\ \hline
asysupstiff   &     0.15              & 0.02     & 8.05  & 15.17   & 5.0  & 4.4   \\ \hline
\end{tabular}
\caption{The ratios $\displaystyle \lambda_2/\lambda_1$, $\displaystyle \lambda_3/\lambda_1$ corresponding to the realistic physical
conditions for the three asy-EOS, the predicted values of PDR, $\displaystyle E_{PDR}$ and GDR, $\displaystyle E_{GDR}$, energy centroids (in MeV), the fraction $\displaystyle f_{PDR}$ exhausted by the PDR in each case. $\displaystyle f_{PDR}^{V}$ reffers to the values obtained from Vlasov calculations.}
\label{taby}
\end{center}
\end{table}
The properties of the region where the total density changes from $\displaystyle \rho_0$ to zero determine the coupling between the core and the excess neutrons.
Therefore we associate the average density of this region with $\displaystyle \rho_i$, obtaining $\displaystyle \rho_i=0.05 fm^{-3}$. The corresponding values of the ratio
$\lambda_2/\lambda_1$, for the three asy-EOS, are reported in Table \ref{taby}.

With these "more realistic" values of the parameters
the PDR energy centroid is found around $8$ MeV for all cases. The EWSR fraction exhausted by PDR is strongly influenced by the density dependence of the
symmetry energy below saturation. 
Values equal to $1.3\%$, $3.3\%$ and  $5.0\%$ are obtained for $\displaystyle f_{PDR}$ when we pass from the asysoft to the superasystiff
parametrization.  
In other words, a stronger coupling between the core and the skin  
reduces the strength of the PDR response \cite{paaRPP2007}, enhancing the GDR contribution. 
Let us mention that in a transport model based on the Vlasov equation,
including both the isovector and the isoscalar channels of the residual interaction, 
it was obtained, for $\displaystyle ^{132}Sn$ \cite{barPRC2013, barEPJD2014}, a PDR peak position around $8$ MeV, 
weakly dependent on the asy-EOS, while the EWSR fraction was $2.4\%$, $4.2\%$ and $4.4\%$, for the three symmetry energy
parametrizations. 
Here the role of the isoscalar component of the residual interaction, which in neutron-rich system may also affect
the isovector response \cite{barPR2005,barPRC2012}, is neglected.
Keeping in mind the crudeness of our assumptions, the agreement between the two models is resonably good,
confirming the clear connection between the behavior of the symmetry energy at quite low densities and the PDR response.

In summary, we introduced in this work schematic models based on separable interactions where the condition of a unique coupling constant for all particle-hole interactions was relaxed. Since the coupling constant for the isovector dipole response can be related to the potential part of the symmetry energy, which is density dependent, the model is well suited to describe situations when part of the nucleons are located in a region at lower density, as in presence of a neutron skin. 
Thus, introducing a density dependent residual interaction for the particles belonging to this region, we find that  
the coherent superposition of particle-hole states 
generate two collective states sharing all the EWSR. For realistic values of the parameters, we reproduce simultaneously the basic experimental features of GDR and PDR, which, within this description, appears as  a collective mode.
 
Finally we further emphasize that the proposed schematic models provide a clear connection between the density dependence of the symmetry energy
and the EWSR exhausted by the PDR. Therefore we consider that precise experimental determinations of the properties of the low energy dipole 
response can settle important constraints on the behavior of the symmetry energy well below saturation. 

For this work V. Baran, A. Croitoru, and A.I. Nicolin were supported by a grant of the Romanian National Authority for Scientific Research, CNCS - UEFISCDI, project number PN-II-ID-PCE-2011-3-0972. A.I. Nicolin was also supported by PN 09370108/2014.

\end{document}